\documentclass[useAMS,usenatbib]{mn2e}
\usepackage{graphicx}
\title[Te~11]{Discovery of an eclipsing dwarf nova in the ancient nova shell Te~11}
\author[Miszalski et al.]{Brent Miszalski,$^{1,2}$\thanks{E-mail: brent@saao.ac.za} P.~A. Woudt,$^{3}$ S.~P. Littlefair,$^{4}$ B. Warner,$^{3}$ H.~M.~J. Boffin,$^{5}$ \newauthor R.~L.~M. Corradi,$^{6,7}$ D. Jones,$^{6,7}$ M. Motsoaledi,$^{3}$ P. Rodr\'iguez-Gil,$^{6,7}$ L. Sabin,$^{8}$ \newauthor and M. Santander-Garc\'ia$^{9,10}$  \\
$^{1}$South African Astronomical Observatory, PO Box 9, Observatory, 7935, South Africa\\
$^{2}$Southern African Large Telescope Foundation, PO Box 9, Observatory, 7935, South Africa\\
$^{3}$Astrophysics, Cosmology and Gravity Centre, Department of Astronomy, University of Cape Town, Private Bag X3,\\ Rondebosch 7701, South Africa\\
$^{4}$Department of Physics and Astronomy, University of Sheffield, Sheffield, S3 7RH, UK \\
$^{5}$European Southern Observatory, Alonso de C\'ordova 3107, Casilla 19001, Santiago, Chile\\
$^{6}$Instituto de Astrof\'isica de Canarias, E-38205 La Laguna, Tenerife, Spain\\
$^{7}$Departamento de Astrof\'isica, Universidad de La Laguna, E-38206 La Laguna, Tenerife, Spain\\
$^{8}$Instituto de Astronom\'ia, Universidad Nacional Aut\'onoma de M\'exico, Apdo. Postal 877, 22860, Ensenada, B. C., Mexico \\
$^{9}$Observatorio Astron\'omico Nacional, Apartado de Correos 112, E-28803, Alcal\'a de Henares, Spain.  \\
$^{10}$Instituto de Ciencia de Materiales de Madrid (CSIC), Sor Juana In\'es de la Cruz, 3, E-28049 Madrid, Spain 
}
\begin{document}

\date{Accepted . Received ; in original form }

\maketitle
\begin{abstract}
   We report on the discovery of an eclipsing dwarf nova (DN) inside the peculiar, bilobed nebula Te~11. Modelling of high-speed photometry of the eclipse finds the accreting white dwarf to have a mass 1.18 M$_\odot$ and temperature 13 kK. The donor spectral type of M2.5 results in a distance of 330 pc, colocated with Barnard's loop at the edge of the Orion-Eridanus superbubble. The perplexing morphology and observed bow shock of the slowly-expanding nebula may be explained by strong interactions with the dense interstellar medium in this region. We match the DN to the historic nova of 483 CE in Orion and postulate that the nebula is the remnant of this eruption. This connection supports the millennia time scale of the post-nova transition from high to low mass-transfer rates. Te~11 constitutes an important benchmark system for CV and nova studies as the only eclipsing binary out of just three DNe with nova shells.
\end{abstract}


\section{Introduction}
\label{sec:intro}
Cataclysmic variables (CVs) are a cyclically erupting group of evolved, interacting close binary stars (Warner 1995). The accumulation of gas on the accreting white dwarf in CVs results in semi-regular thermonuclear runaway explosions commonly referred to as novae. Following such an explosion, the surface of the white dwarf is heated and the white dwarf accretes gas from its companion at an increased mass transfer rate due to strong irradiation effects on the donor star. This phase is expected to last a few centuries (Shara et al. 1986). Post-nova mass transfer rates reduce over time, allowing the accretion disc in the binary to experience standard disc instabilities (Schreiber et al. 2000; Schreiber \& G\"ansicke 2001; Lasota 2001), resulting in regular day-to-week long outbursts (termed dwarf novae, DNe).
In recent years much interest has been focused towards finding DNe with faint nova shells left over from recent ($\sim$1000 yr old) eruptions, allowing for a holistic view of the evolutionary cycle of CVs as a group of cyclically erupting binaries. To date, only 2 firm cases have been identified, Z Cam (Shara et al. 2007, 2012a) and AT Cnc (Shara et al. 2012b); both are Z Cam-type CVs -- systems with the highest mass transfer rates amongst DNe. The DN V1363 Cyg has a candidate nova shell (Sahman et al. 2015) and these shells are exceedingly rare (Schmidtobreick et al. 2015; Sahman et al. 2015).

Te~11 (Table \ref{tab:summary}) was identified as a possible planetary nebula (PN) candidate by Jacoby et al. (2010). Its peculiar morphology (Sect. \ref{sec:morph}) motivated its inclusion in a photometric monitoring program to identify binary central stars responsible for shaping PNe (Miszalski et al. 2011). An eclipsing lightcurve with an orbital period of 0.12 days was obtained that is not out of place for binary central stars of PNe. Here we report on new observations of the nebula Te~11 and its central star that has since been classified as a CV (Thorstensen \& Skinner 2012; Drake et al. 2014). This paper is structured as follows. Section \ref{sec:phot} presents new observations of the nucleus, focusing on modelling of high-speed photometry to derive the binary star parameters. Images and spectroscopy of the nebula are presented in Sect. \ref{sec:bilobal} that suggest the nebula is an ancient nova shell rather than a PN. We conclude in Sect. \ref{sec:conc}.
\begin{table}
   \centering
   \caption{Derived parameters of the DN nucleus of Te~11.}
   \label{tab:summary}
   \begin{tabular}{lr}
\hline
Right ascension & 05 45 58.2 \\ 
Declination & $+$02 21 06 \\ 
Distance (pc) & $330\pm50$ \\ 
Orbital period (days) & $0.1209716\pm0.0000002$ \\ 
Mid-eclipse time (HJD$_0$) & 2456656.39483\\
Mass ratio $q$ & $0.236\pm0.006$ \\ 
WD mass $M_w$ (M$_\odot$)& 1.18$^{+0.07}_{-0.15}$ \\ 
WD radius $R_w$ (R$_\odot$) & 0.006$^{+0.002}_{-0.001}$  \\ 
WD temperature $T_w$ (K) & 13000$^{+5000}_{-2500}$ \\ 
Donor mass $M_d$ (M$_\odot$)& $0.28\pm0.03$ \\ 
Donor radius $R_d$ (R$_\odot$) & $0.31\pm0.01$ \\ 
Donor temperature $T_d$ (K) & $3400\pm100$ \\ 
Inclination $i$ (deg) & $85.4\pm0.4$ \\ 
Orbital separation $a$ (R$_\odot$) & 1.17$^{+0.02}_{-0.05}$ \\ 
\hline
   \end{tabular}
\end{table}

\section{An eclipsing dwarf nova}
\label{sec:phot}
\subsection{Classification and high-speed photometry}

\begin{figure}
   \begin{center}
      \includegraphics[scale=0.33,angle=270]{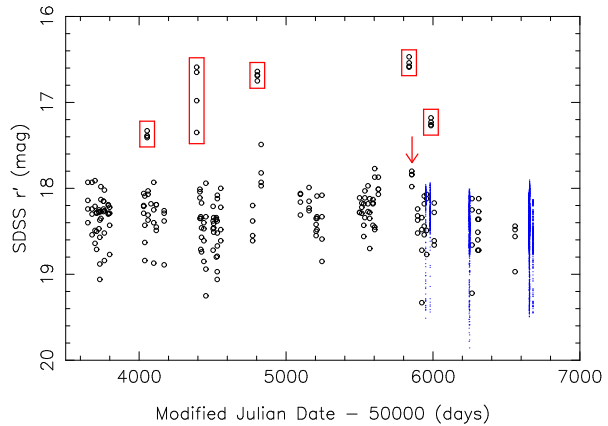}
   \end{center}
   \caption{CRTS ($V$ filter, black circles) and SHOC (clear filter, blue points) photometry of the central star of Te~11. DN outbursts (red squares) and deep eclipses are evident. An arrow indicates the epoch of the Drake et al. (2014) spectrum.}
   \label{fig:phot}
\end{figure}

\begin{figure}
   \begin{flushleft}
      \vspace{-0.7cm}
      \hspace{-7mm}
      \includegraphics[scale=0.45]{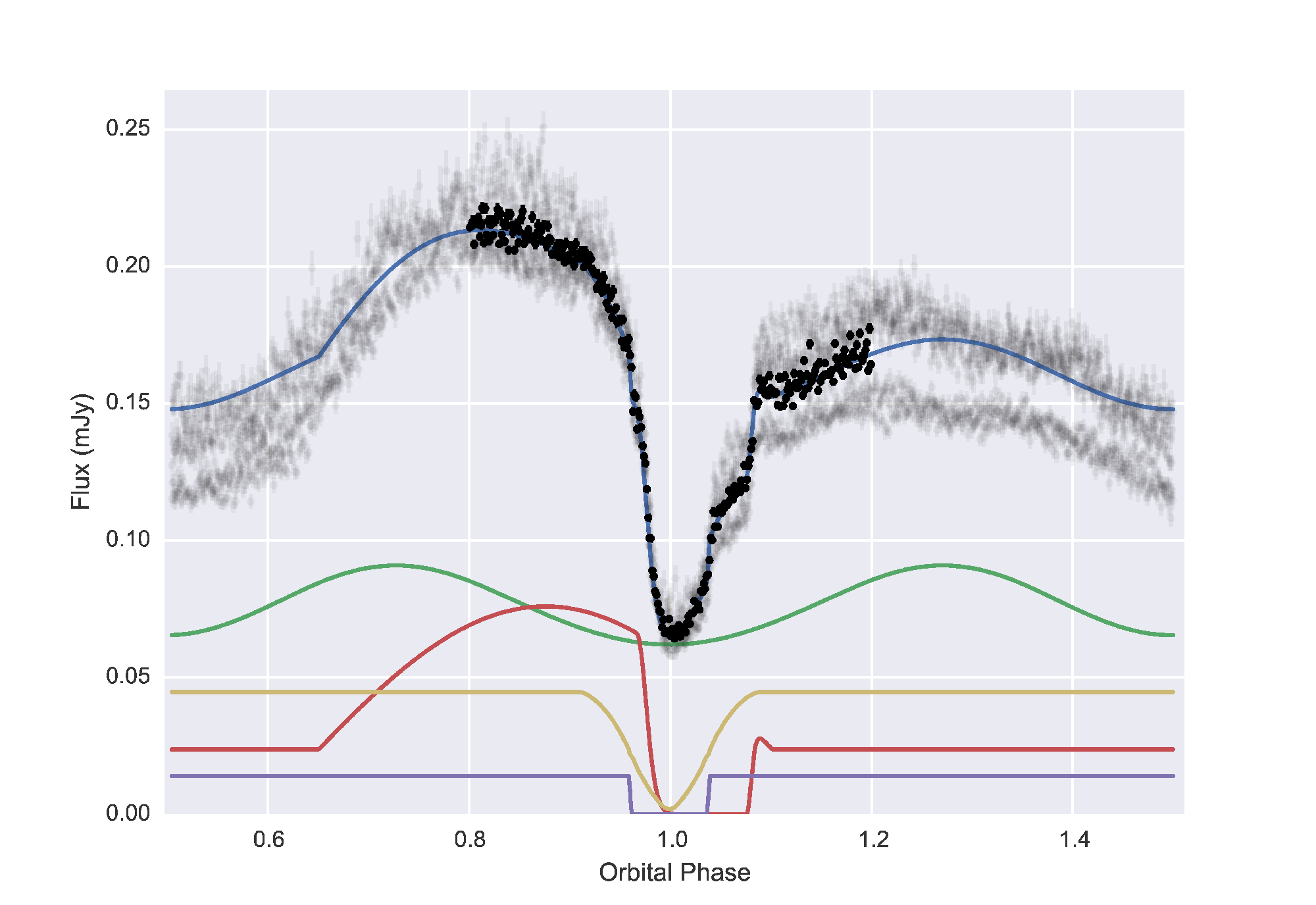}
   \end{flushleft}
   \caption{Model best fit (blue line) of the SHOC photometry (grey points, 20 s time-resolution) phased on the orbital period. The eclipse is depicted with binned data points (black points) and is an admixture of different components including the WD (purple line), accretion disc (yellow line), the disc bright spot (red line) and ellipsoidal variability (green line).}
   \label{fig:fit}
\end{figure}

The Catalina Real-Time Transient Survey (CRTS, Drake et al. 2009) has recorded five outbursts in the nucleus of Te~11 over 10 years (Fig. \ref{fig:phot}). The outburst behaviour prompted us to obtain multi-epoch high-speed photometry of the transient source since Jan 2012 (Figures \ref{fig:phot} and \ref{fig:fit}). Table \ref{tab:photlog} summarises the observations of 20 s time resolution taken with the SHOC camera (Coppejans et al. 2013) on the SAAO 1.9-m telescope in white light (unfiltered). Magnitudes were converted to SDSS $r'$ as described in Woudt et al. (2012) and are accurate to $\sim$0.1 mag. The converted CRTS data are thus also corrected for the nebular contamination which we assume is constant. The phased lightcurve shape and the spectrum taken on 23 Oct 2011 by Drake et al. (2014), 20 days after a recorded outburst, firmly classify the nucleus as a non-magnetic DN (Warner 1995).

\begin{table}
   \centering
   \caption{Observation log of high-speed photometry of the DN nucleus of Te~11. The magnitude range specifies the brightest out-of-eclipse measurement to the measurements at mid-eclipse.}
   \label{tab:photlog}
   \begin{tabular}{lllll}
      \hline
Run & Date & HJD & Length & SDSS $r'$   \\
    &              & (first obs.) & (h) & (mag) \\
\hline
S8355 & 2013 Dec 29 & 2456656.33831 & 5.75 & 18.1 - 19.4 \\
S8356 & 2013 Dec 30 & 2456657.31034 & 5.73 & 18.0 - 19.5 \\
S8358 & 2013 Dec 31 & 2566658.30586 & 5.38 & 17.9 - 19.4 \\
S8360 & 2014 Jan 23 & 2456681.29749 & 2.50 & 18.1 - 19.4 \\
\hline
   \end{tabular}
\end{table}

The high time resolution photometry allows us to analyse the eclipse profile in greater detail. The light-curves were phased according to our derived ephemeris (Table \ref{tab:summary}) and then binned into 290 data points between orbital phases 0.8 and 1.2 to produce an average light curve (Fig \ref{fig:fit}). Sharp steps in the light curves represent the ingress and egress of the WD behind the donor. The largest feature is the eclipse of the bright spot, where the gas stream hits the outer edge of the accretion disc. The presence of a bright spot confirms ongoing accretion, validating our assumption that the donor fills its Roche Lobe. The binned lightcurve is fitted with a geometric model including a limb-darkened WD and a bright spot modelled as a linear strip passing through the intersection of the gas stream and accretion disc. 

The following section gives details of the modelling procedure and the model results are combined with a theoretical WD mass-radius relationship to obtain a full solution for the binary parameters (Table \ref{tab:summary}). The individual lightcurves shown in Fig \ref{fig:phot} show some variations in the strength of the emission from the bright spot and the accretion disc. This is not unexpected (e.g. McAllister et al. 2015) and does not affect the main results of the modelling. An important ingredient in the modelling is the previously unknown spectral type of the donor star.

\begin{figure}
   \begin{center}
      \includegraphics[scale=0.5,angle=270]{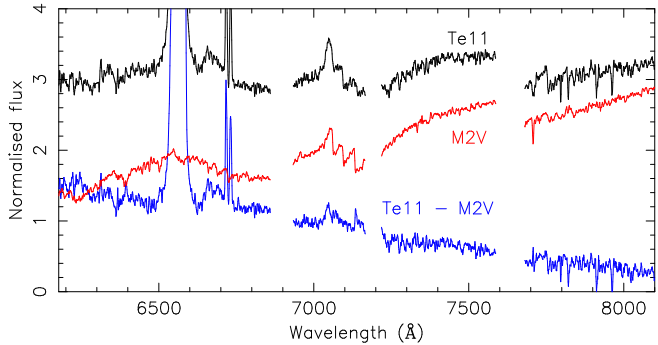}
   \end{center}
   \caption{SALT RSS spectrum of Te~11 (black) after dereddening and smoothing with a 3$\times$3 boxcar filter. The M2V template spectrum (red, Bochanski et al. 2007) was subtracted from the Te~11 spectrum resulting in a smooth blue continuum with weak double-peaked profiles from He~I 6678 and 7065, all originating from the accretion disc. The absorption features near 7900 \AA\ are sky subtraction residuals.}
   \label{fig:salt}
\end{figure}

To determine the spectral type we observed Te~11 with the Robert Stobie Spectrograph (RSS; Burgh et al. 2003; Kobulnicky et al. 2003) on the Southern African Large Telescope (SALT; Buckley, Swart \& Meiring 2006; O'Donoghue et al. 2006) on 7 March 2015. A 40 min longslit spectrum was taken with the PG900 grating and 1.5\arcsec\ slit covering wavelengths 6176-9142 \AA\ at 5.7\AA\ resolution (FWHM). After basic processing (Crawford et al. 2010) the data were cleaned of cosmic ray events (van Dokkum 2001) and reduced with \textsc{iraf}. Wavelengths redder than 8100 \AA\ were significantly affected by detector fringing and were disregarded. To classify the spectral type we follow the method described by Thorstensen et al. (2010). M-dwarf template spectra (Bochanski et al. 2007) were individually scaled to best match the features around 7100 \AA\ in the dereddened Te~11 spectrum before they were subtracted from the spectrum. The only resulting spectra without strong residues of the M-dwarf template spectra were those from the M2 and M3 templates. Figure \ref{fig:salt} shows the Te~11 spectrum and the residual after subtracting the M2 template. A relatively smooth spectrum remains with a blue continuum that is dominated by light from the accretion disc. We therefore assign a spectral type of M2.5$\pm$0.5, corresponding to an effective temperature of 3400$\pm$100 K (Rajpurohit et al. 2013).

\subsection{A parameterised model of the eclipse}
\label{sec:model}
As the gas stream follows a ballistic trajectory from the donor to the edge of the accretion disc, the timing of the gas stream provides information on the mass ratio of the system. By fitting a physical model of the binary system to the eclipse it is possible to determine the binary parameters (Littlefair et al. 2014). Extracting the system parameters from the eclipse light curve depends on four assumptions: that the bright spot lies on the ballistic trajectory from the secondary star; that the WD is accurately described by a theoretical mass-radius relation; that the whole of the WD is visible out of eclipse and that the donor star fills its Roche lobe. We cannot be sure that all of these assumptions hold for every system, but we point out here that dynamical mass determinations of CVs agree with those found via eclipse fitting (Tulloch et al. 2009; Copperwheat et al. 2010, 2012; Savoury et al. 2012).

The model we fit is described in detail by Savoury et al. (2011). To summarise, the free parameters are:
\begin{itemize}
\item	the mass ratio, $q$;
\item	the WD eclipse phase full-width at half depth, $\Delta \Phi_{1/2}$;
\item the (linear) WD limb-darkening parameter, $U_w$;
\item	the WD radius, scaled to the orbital separation, $R_w/a$;
\item	seven parameters describing the emission from the bright-spot;
\item	a disc exponent, specifying the power law of the radial intensity distribution of the disc;
\item	a phase offset;
\item	the flux contributions of the WD, donor star, accretion disc and bright spot.
\end{itemize}

\begin{figure*}
   \begin{center}
      \includegraphics[scale=0.5]{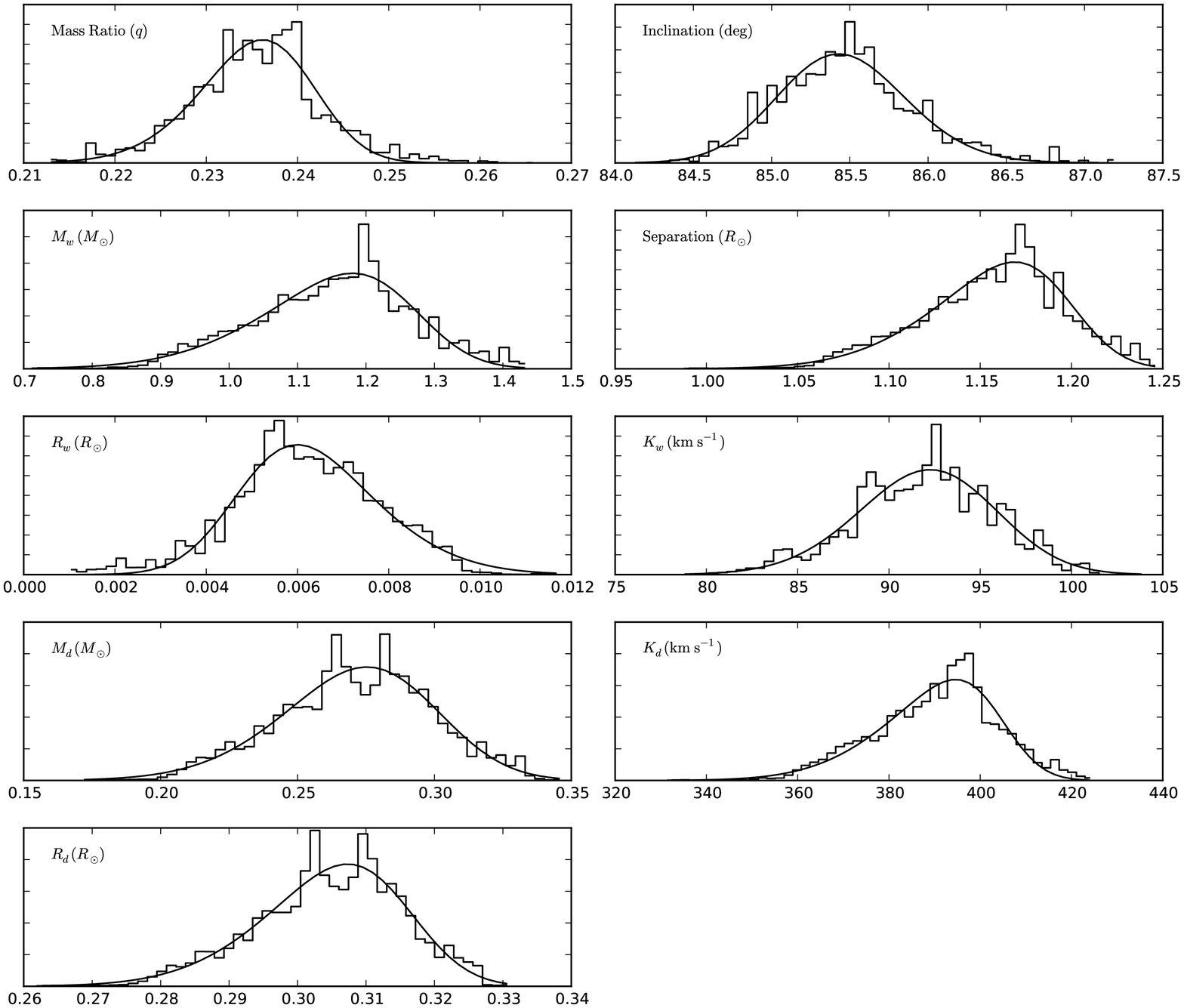}
      \vspace{-0.5cm}
   \end{center}
   \caption{Posterior probability distributions for the system parameters quoted in Table \ref{tab:summary}. Solid lines represent the best fit of a skewed Gaussian to the histograms.}
   \label{fig:posterior}
\end{figure*}

The data cannot constrain the WD limb darkening parameter. Instead, we fix this parameter at $U_w=0.35$. Changing the limb darkening parameter across all plausible values does not affect the results. We draw samples from the posterior distributions of our remaining parameters by a Markov-Chain Monte Carlo (MCMC) procedure. Posterior probability distributions are estimated using an affine-invariant ensemble sampler (Foreman-Mackey et al. 2013). Uninformative priors were used for all parameters. The MCMC chains consist of a total of 400,000 steps of which 200,000 were discarded as burn-in. We visually examined the chains for convergence, requiring that the means and root-mean-square values for all parameters showed no long term trends. The Chi-squared of the most probable model was 1200 with 270 degrees of freedom, a consequence of the large variability in total flux between individual eclipses. Our MCMC chains provide an estimate of the joint posterior probability distributions of ($q$, $\Delta \Phi_{1/2}$, $R_w/a$). At each step in the chain we can calculate the system parameters from these three values, Kepler's third law, the orbital period, and a WD mass-radius relationship (corrected to our derived WD temperature, see Sect. \ref{sec:dist}).\footnote{This is an iterative procedure where the system parameters are first estimated using a guess of the WD temperature.} We favour mass-radius relationships with thick hydrogen layers (Wood 1995), but these only extend to 1.0 M$_\odot$. Above these masses we adopt mass-radius relationships with thinner hydrogen layers (Panei et al. 2000). The results are posterior probability distributions for the system parameters quoted in Table \ref{tab:summary} and depicted in Figure \ref{fig:posterior}, with the exceptions of the distance and WD temperature, which are estimated in the next section. 

\subsection{Distance and WD temperature}
\label{sec:dist}
The donor mass, radius and effective temperature are fully consistent with predictions for main sequence stars (Dotter et al. 2008). Our modelling gives a measure of the observed $r'$-band flux and radius of the donor star. The distance is obtained by comparing these measures to the predicted $r'$-band surface brightness of the donor. We use the measured donor spectral type of M2.5$\pm$0.5 to estimate a donor star colour of $(g'-i')$ = 2.53$\pm$0.15 (Covey et al. 2007). This colour is used to look up the interferometric surface brightness of the donor (Boyajian et al. 2014) in the $g'$-band. The surface brightness is given in terms of the zero-magnitude angular diameter, log$_{10}$ $\theta_{(g'=0)}$. From the definition of this quantity, we can write that the distance to the system is given by
\[\log_{10} d = \log_{10} 2R + 0.2 r'- 0.2 A_{r'} - \log_{10}\theta_{(g'=0)} + 0.2 (g'-r')\]
where $R$ is the donor star radius, measured in AU, $r'$ is the observed $r'$-band magnitude of the donor, and $A_{r’}$ is the extinction towards the system in the $r'$-band. We estimate $A_{r’}$ from the measured extinction of $E(B-V) = 0.38$ (Sect. \ref{sec:neb}), which we convert to the $r'$-band extinction using empirical relationships (Yuan et al. 2013). The donor's $(g'-r')$ colour is estimated from the observed spectral type (Covey et al. 2007). We perform this calculation for 10,000 random samples from the posterior distributions for the donor flux and radius and find a distance of $330\pm50$ pc.

The derived distance agrees well with an independent estimate of $d$ = $420\pm100$ pc calculated using the recalibrated relationship between the absolute magnitude at maximum brightness $M_{V,\mathrm{max}}$, inclination and orbital period (Warner 1987; Patterson 2011). We obtained a value of $M_{V,\mathrm{max}}$ = 7.38 mag, including a correction of 2.51 mag to account for the 85 degree inclination of Te~11, while the apparent magnitude equivalent corrected for reddening is $m_{V,\mathrm{max}}$ = 15.5 mag.

With a known distance and reddening, we can convert the modelled white dwarf flux into an absolute $r'$-band magnitude. This can be directly compared to the predictions of models (Bergeron et al. 1995). We draw 5000 random samples from the posterior distributions of the white dwarf mass, radius and $r'$-band flux. For each sample we use a Brent root-finding algorithm to find the white dwarf temperature which best satisfies these three constraints. We find a white dwarf temperature of $T_w$ = 13000$^{+5000}_{-2500}$ K. The reliability of this temperature estimate depends on several factors; the absolute photometric calibration of the lightcurve, the robustness of measuring the white dwarf flux from the eclipse lightcurve and the robustness of the distance. In cases where eclipses in multiple photometric bands are used, and the time resolution of the lightcurve is high enough to fully resolve the eclipse of the white dwarf, temperatures derived this way are accurate to $\sim$1000 K (Littlefair et al. 2008). In our case we only have a single band, and the time resolution of the lightcurve is too low to resolve the white dwarf eclipse, making it difficult to robustly estimate the white dwarf flux. Systematic errors arising from these limitations will likely be of similar size to the formal errors quoted above, but future observations with high cadence and in several wavebands will give a more reliable white dwarf temperature.

\section{The bilobal nebula}
\label{sec:bilobal}
\subsection{Morphology and expansion velocity}
\label{sec:morph}
\begin{figure*}
   \begin{center}
      \includegraphics[scale=0.25]{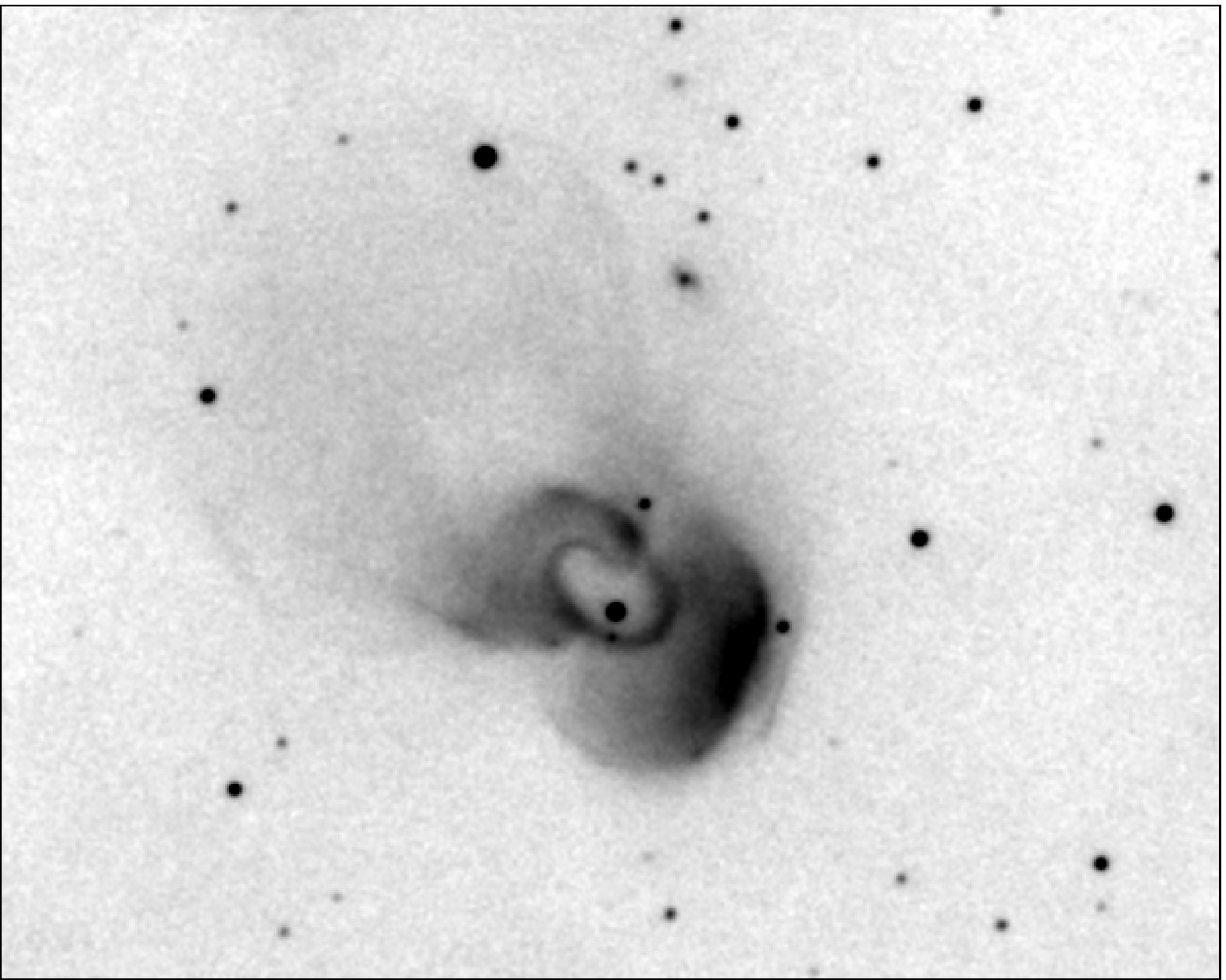}
      \includegraphics[scale=0.25]{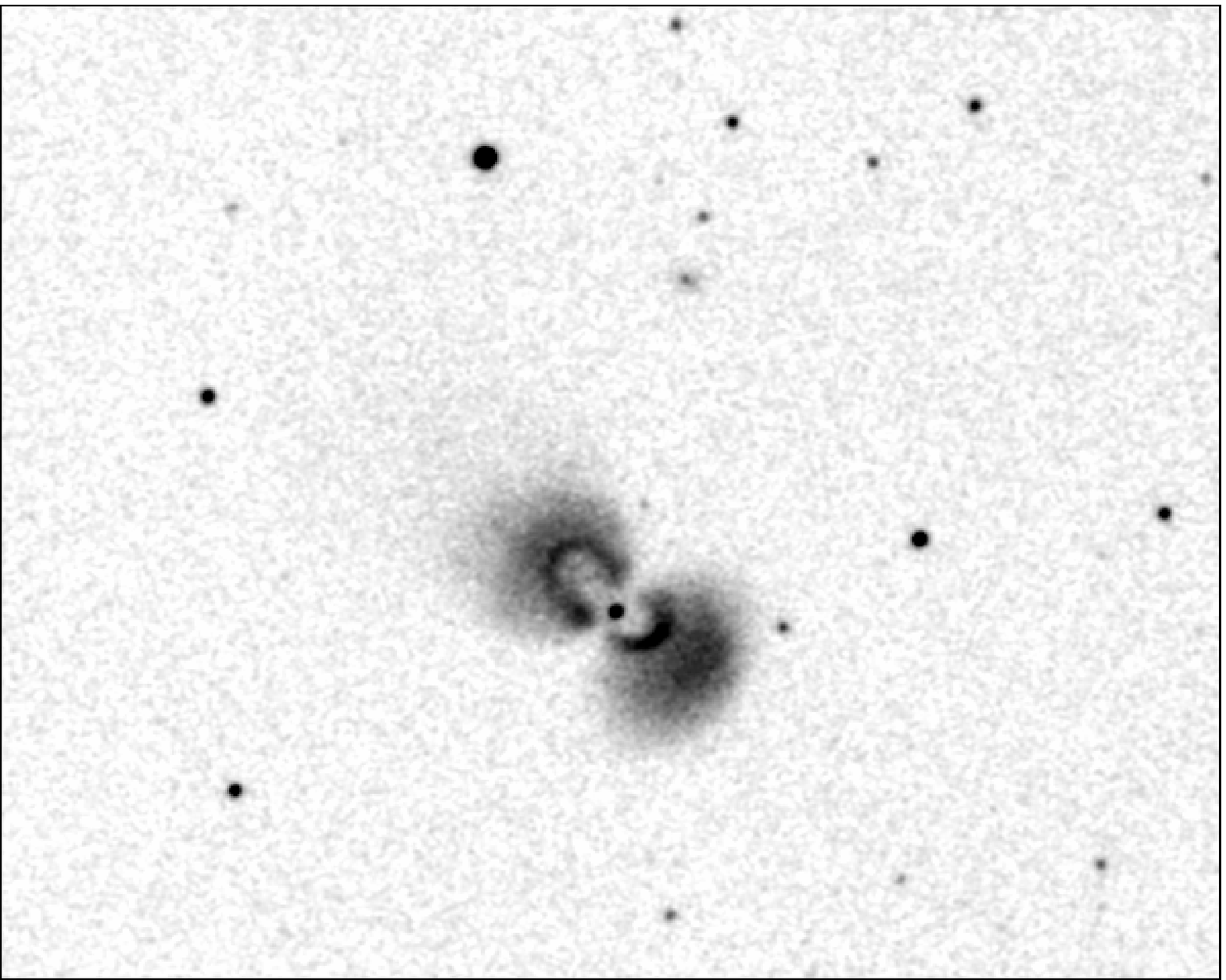}
      \includegraphics[scale=0.24791]{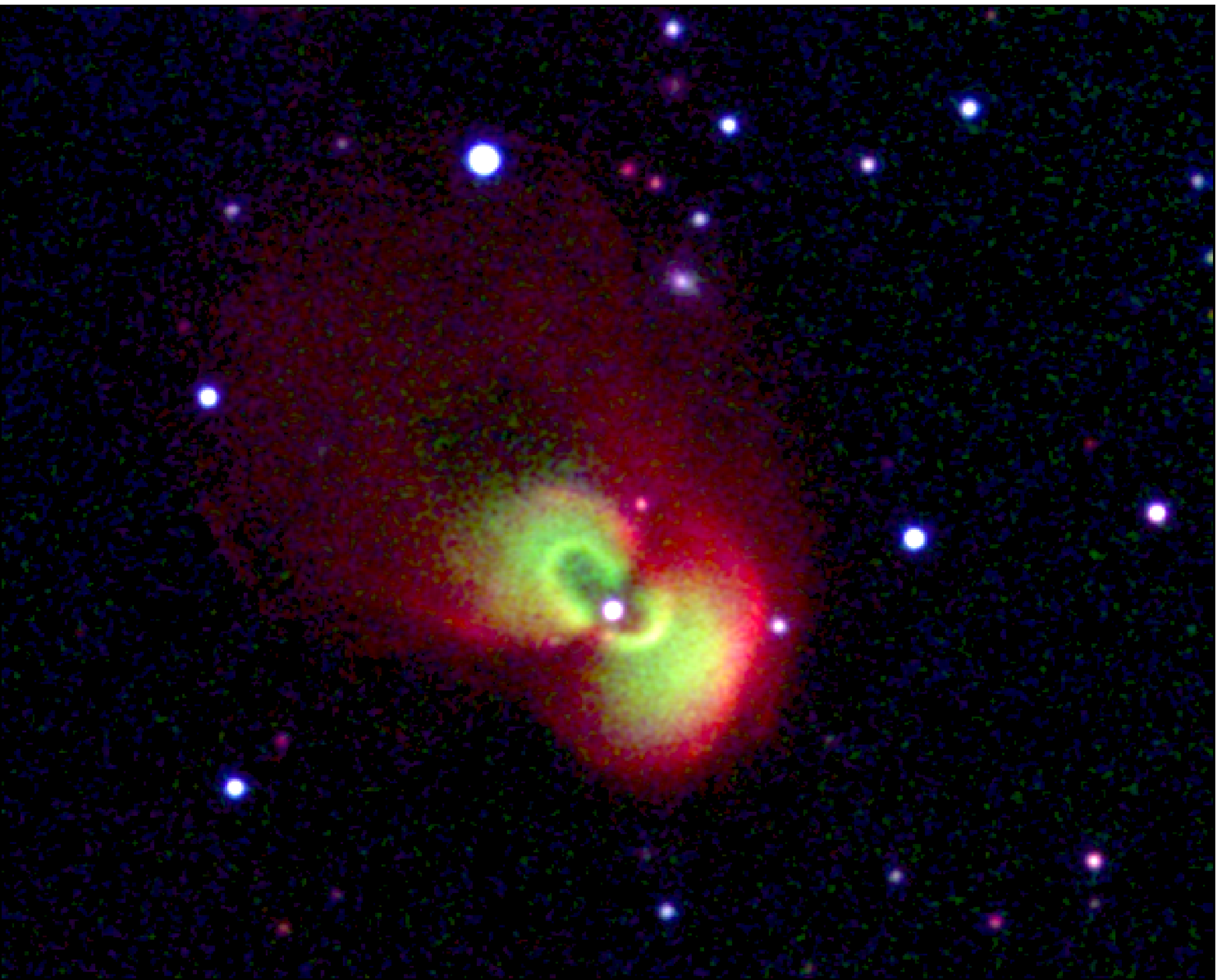}
   \end{center}
   \caption{VLT FORS2 images of Te~11 in H$\alpha$+[N II] (left) and [O III] (middle) filters. The colour-composite image is made from H$\alpha$+[N II] (red), [O III] (green) and SDSS $g'$ (blue, Alam et al. 2015). All images measure 2.5$\times$2.0 arcmin$^2$ with North up and East to the left.}
   \label{fig:neb}
\end{figure*}

Figure \ref{fig:neb} depicts VLT FORS2 (Appenzeller et al. 1998) images of the nebula taken with H\_Alpha$+$83 (H$\alpha$+[N II]) and OIII$+$50 ([O III]) filters on 12 February 2012. We measured a nebular radius in [O III] of 17.6 arcsec (35.3 arcsec diameter) and in H$\alpha$+[N II] the major axis (including dispersed material) measures 87.9 arcsec. At the 330 pc distance the radius and major axis extent measures 0.03 pc and 0.14 pc, respectively. The central star position is centered with respect to [O III] emission, but is offset in H$\alpha$, presumably due to the ongoing interstellar medium (ISM) interaction (Wareing et al. 2007). The bow shock lacks [O III] emission, indicating Te~11 is moving slowly with respect to the ISM (e.g. Shull \& McKee 1979). This rules out shock-excitation due to fast motion through the ISM as the excitation mechanism for [O III], unlike that observed in BZ Cam (Hollis et al. 1992). Te~11 is not detected at 22 microns by the Wide-field Infrared Survey Explorer mission (Wright et al. 2010).

Spatially-resolved, longslit emission line spectra of Te~11 were acquired with the second Manchester Echelle Spectrometer mounted on the 2.1-m San Pedro M\'artir Telescope (MES-SPM, Meaburn et al. 2003). On 23 Nov 2009, a 20 min exposure was acquired with a narrowband filter to isolate the H$\alpha$ and [N~II] 6548, 6583 \AA\ emission lines with a spatial scale of 0.61 arcsec pix$^{-1}$ and a spectral scale of 4.4 km s$^{-1}$ pix$^{-1}$.  On 11 Dec 2010, a 30 min exposure was acquired with a narrowband [O III] filter in order to isolate the 5007 \AA\ emission line with a spatial scale of 0.36 arcsec pix$^{-1}$ and a spectral scale of 2.7 km s$^{-1}$. In both cases the longslit was oriented along the major axis of the nebula.

\begin{figure}
   \begin{center}
      \includegraphics[angle=270,scale=0.65]{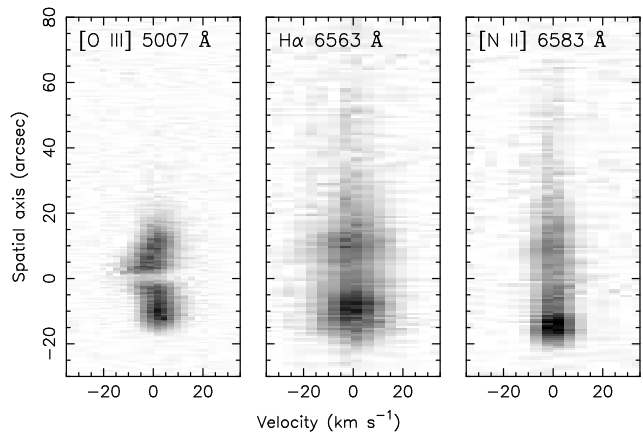}
   \end{center}
   \caption{Position-velocity arrays of MES observations of Te~11 in three nebular emission lines. The spatial axis increases from the south-west (bottom of figure) towards the north-east (top of figure) along the major axis of the nebula. (linear greyscale).}
   \label{fig:echelle}
\end{figure}

Figure \ref{fig:echelle} shows the observed emission line profiles of [O III], H$\alpha$ and [N II] after sky subtraction. The spatial axis is defined such that the central star is at zero and the velocity axis has been corrected for the measured heliocentric systemic velocity of 28$\pm$3 km s$^{-1}$. Apart from the thermally broadened complex nature of the H$\alpha$ profile, we found the full-width at half-maximum (FWHM) of the position-velocity arrays to be consistent with a very low expansion velocity of $2V_\mathrm{exp}<10$ km s$^{-1}$. A bilobed nebula geometry would be consistent with the observations, however insufficient information is available to place any further constraints on the geometry via spatio-kinematic modelling.

\subsection{Nebular spectrum and ionized mass}
\label{sec:neb}
\begin{figure}
   \begin{center}
      \includegraphics[scale=0.35,angle=270]{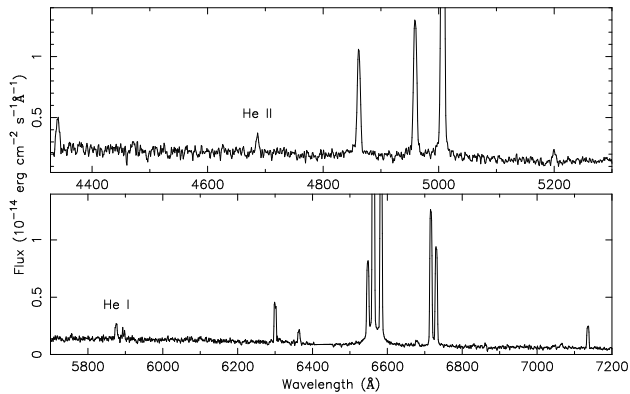}
   \end{center}
   \caption{SALT RSS nebular spectrum of Te~11.}
   \label{fig:saltneb}
\end{figure}

Another SALT RSS longslit spectrum of 30 min duration was taken of Te~11 on 28 Oct 2012 with the PG900 grating and 1.5\arcsec\ slit covering wavelengths 4335-7413 \AA\ at 6.2 \AA\ resolution (FWHM). The longslit was oriented along the major axis and after basic processing (Crawford et al. 2010) the data were cleaned of cosmic ray events (van Dokkum 2001) and reduced with \textsc{iraf}. The spectrum was extracted by averaging over the spatial extent of the nebula and a relative flux calibration was applied. Figure \ref{fig:saltneb} shows the spectrum and Tab. \ref{tab:lines} gives the emission line intensities. The reddening from the Balmer decrement is $E(B-V)=0.32$ mag or $A_V=1.0$ mag. The electron density measured from the ratio [S II] 6717/6730 is constant across the major axis at 62 cm$^{-3}$. An integrated SDSS $r'$ flux of the nebula was determined from aperture photometry of the flux calibrated $r'$ image (Alam et al. 2015) with an aperture of diameter 36.8 arcsec (centered near the central star). We then convolved the SALT nebular spectrum with the SDSS $r'$ filter using the \textsc{synphot} software, scaling the spectrum so that its $r'$ flux matches the measured $r'$ flux from aperture photometry. An H$\alpha$ flux of 2.4$\times$10$^{-13}$ erg s$^{-1}$ cm$^{-2}$ was measured from the calibrated spectrum. Assuming a gas temperature of 10000 K and case B conditions, we used the equation in Corradi et al. 2015 to obtain a total ionized mass of hydrogen of $\sim$2$\times$10$^{-4}$ M$_\odot$. Given the aperture photometry was restricted to the main part of the nebula, the ionized mass for the whole nebula will be slightly higher.

\begin{table}
   \centering
   \caption{Observed emission line intensities of Te~11. The values represent an average of the total emission along the major axis, normalised to H$\beta$=100, and have not been corrected for interstellar extinction.}
   \label{tab:lines}
   \begin{tabular}{ll}
      Line & Intensity\\
      \hline
     {}H$\gamma$ 4340 & 23.8\\ 
     {}He~II $\lambda$4686 & 15.9\\
     {}H$\beta$ 4861 & 100.0\\
     {}[O~III] 4959 & 135.0\\
     {}[O~III] 5007 & 414.6\\
     {}[N~I] 5200 & 11.3\\
     {}He~I 5876 & 17.0\\
     {}[O~I] 6300 & 41.1\\
     {}[O~I] 6363 & 12.3\\
     {}[N~II] 6548 & 100.0\\
     {}H$\alpha$ 6563 & 404.9 \\
     {}[N~II] 6584 & 243.7\\
     {}He~I 6678 & 6.7\\
     {}[S~II] 6717 & 141.7\\
     {}[S~II] 6730 & 103.9\\
     {}[Ar~III] 7135 & 22.3\\
      \hline
   \end{tabular}
\end{table}

\subsection{Te 11 as an ancient nova shell}

The location of Te~11 (Tab. \ref{tab:summary}) is close to the easternmost star in Orion's Belt (Alnitak; $\alpha=$05h 41m, $\delta=-$01$^\circ$ 57$'$). Significantly, Alnitak is the principal star in the ancient Chinese Lunar Mansion 21 (called Shen or Three Stars). A search through the recorded oriental novae (Xu et al. 2000) reveals an obvious candidate, the nova of 16 Nov--14 Dec 483 CE, which was described as ``Emperor Xisosen of Wei, 7th year of the Taihe reign period, 10th month. There was a guest star east of Shen as large as a peck measure and like a fuzzy star''. The reference to a ``fuzzy star'' might be thought to indicate a comet, but Nickiforov (2010) finds that many of the “fuzzy star” descriptions cannot refer to comets and must be novae or supernovae. Assuming an absolute magnitude of $M_V=-8$ mag (Della Valle 1991) and including the $A_V=1.0$ mag of extinction, a nova at the 330 pc distance of Te~11 would have $V=0.6$ mag. Some novae reach a maximum $M_V=-10$ mag, which could have made a nova at this distance appear as bright as Jupiter.

Could the Te~11 nebula be the remnant of this explosion 1532 years ago? The appearance of a nova inside a PN would be an oddity, but is not unprecedented (Wesson et al. 2008; Rodr{\'{\i}}guez-Gil et al. 2010). At face value Te~11 resembles a PN, presenting a bilobed nebula interacting strongly with the ISM (Wareing et al. 2007), however there are several arguments against a PN classification. The ionized mass of the nebula of $\sim$2$\times$10$^{-4}$ M$_\odot$ is far too low for a PN (Meatheringham et al. 1988). Also atypical of PNe is the absence of 22 micron emission (c.f. Chu et al. 2009; Mizuno et al. 2010) and the nebular expansion velocity of $2V_\mathrm{exp}<10$ km s$^{-1}$ (c.f. Richer et al. 2010). Furthermore, the overall nebular spectrum shows considerable similarity with the nova shell of AT Cnc (Shara et al. 2012b) and the `jet-like' feature of GK Per (Shara et al. 2012c). The presence of He II requires a photoionizing source with T $\ga$ 60 kK, much hotter than the model-constrained WD temperature of 13 kK (a much hotter WD would also switch off the DN outbursts). One potential explanation could be that a hotter WD companion to the DN were present (i.e. an unresolved triple), but there is no evidence of one in our observations and this can probably be excluded. Perhaps more likely is that the present-day weak He II emission is a residue of a nova explosion having since faded considerably (Shore et al. 1996). Identifying the true excitation mechanism of the nebula will require further detailed observations.

\subsection{An unusual neighbourhood: Barnard's Loop}
While the measured ionized mass of the nebula is consistent with a nova shell interpretation (Shara et al. 2012a), the current expansion velocity is anomalously low for a historic nova shell (Shara et al. 2012a). This might be explained by ISM interaction, with a bow shock evident in the South-West and dispersed gas in the North-East of the nebula (Fig. \ref{fig:neb}). Indeed, at a distance of 330$\pm$50 pc, Te~11 is co-spatial with the Orion-Eridanus bubble towards the edge in Barnard's Loop (Madsen et al. 2006; O'Dell et al. 2011). This is a region with enhanced gas densities and a nova blast wave propagating through a denser ISM will therefore be more compact, and will have reduced shock velocities (see eqns. 1 and 2 of Contini \& Prialnik 1997). A factor of 100 increase in ambient densities will lead to a factor of 2.5 reduction in the extent of the nebula and the shock velocities of the blast wave, in comparison to blast wave propagation at average ISM densities. This could explain the smaller size of the shell of Te~11 and the small measured expansion velocity, compared with Z Cam and AT Cnc. Alternatively, Te~11 might conceivably be the result of an interaction between the dense ISM and weak winds from the DN, but detailed simulations are needed to assess whether this is feasible. 

\vspace{-0.7cm}

\section{Conclusions}
\label{sec:conc}
We have presented further observations of the peculiar nebula Te~11 (Jacoby et al. 2010) and its binary central star with an orbital period of 0.12 d (Miszalski et al. 2011). A DN classification for the central star was determined based on the presence of outbursts in the longterm CRTS lightcurve, the shape of the eclipsing lightcurve and the non-magnetic CV spectrum. High-speed photometry of the DN nucleus obtained at SAAO reveals the various components of the system including the WD, the accretion disc, the disc bright spot and ellipsoidal variability. A model fit was applied to these components using the model described by Savoury et al. (2012), allowing for a first estimate of the binary parameters. In addition to the lightcurve, the model was constrained by the spectral type of the donor determined from SALT spectroscopy. The high inclination (85.4 deg) binary consists of a cool 13 kK WD with mass 1.18 M$_\odot$ and radius 0.006 R$_\odot$ together with a 3400 K M-dwarf donor with mass 0.28 M$_\odot$ and radius 0.31 R$_\odot$. The modelling also allowed for the distance to be determined, $330\pm50$ pc, consistent with an independent estimate following the method of Warner (1987).

Narrow-band images of the nebula in the emission lines of [O~III] and H$\alpha$+[N~II] were presented, revealing the peculiar morphology. While the nebula had been previously suspected to be a PN, the ionized mass of 2$\times$10$^{-4}$ M$_\odot$ is too low for a PN classification. Furthermore, the nebula lacks a detection at 22 microns and has a very low expansion velocity ($2V_\mathrm{exp}<10$ km s$^{-1}$), both of which are atypical for PNe. Rather than a PN, we postulate that the nebula is the remnant of an ancient nova eruption in 483 CE that we connect to Te~11. The dense ISM surrounding Te~11 may help explain the current low expansion velocity of the nova shell. Further detailed observations and simulations of the Te~11 nebula are required to understand its formation and especially its excitation source. Its bilobal nature seems consistent with the increasing trend for novae to exhibit bipolar outflows (e.g. Woudt et al. 2009; Schaefer et al. 2014). The unique eclipsing nature of Te~11, combined with the relatively bright and nearby nova shell, makes it particularly valuable for detailed examination of all physical parameters to inform models of the evolution of CVs, nova shells and their interaction with the ISM.

The massive white dwarf of Te~11 ($M_w$ = 1.18 M$_\odot$) could suggest that it is a frequent recurrent nova, with explosions every few hundred years (Yaron et al. 2005). This time scale is poorly constrained for Te~11, although the $\sim$1500 yr time scale from last recorded nova eruption to the present state of regular DN (disc instability) outbursts in Te~11 is consistent with the evolution/variation of mass transfer rate in short-period CVs through a nova cycle (e.g. BK Lyn, Patterson et al. 2013). A higher angular resolution image of the nebula of Te~11 might show evidence of multiple nova eruptions and the interaction of more recently ejected material with older ejecta, as was seen in the recent eruption of the recurrent nova T Pyx (Shara et al. 2015). In this context the buildup of material seems to be a plausible explanation for the agreement between the ionized masses of Te~11 (2$\times$10$^{-4}$ M$_\odot$) and T Pyx (7$\times$10$^{-4}$ M$_\odot$, Shara et al. 2015).

The discovery of a DN inside Te~11 provides a potential new strategy to ameliorate the paucity of known ancient nova shells. Follow-up observations of faint central stars of existing PNe, most of which have not been studied in detail, could form part of such a strategy. After all, there remain many ancient novae that lack associations. The focus of current transient surveys is mostly at high Galactic latitudes where there are very few PNe. Future transient surveys that cover the Galactic Plane would be ideally placed to identify Te~11-like objects amongst known PNe. Similarly, as we have done here, existing transient surveys would benefit from cross-checking with newly discovered high-latitude PNe (e.g. Jacoby et al. 2010). 
\vspace{-0.7cm}
\section{Acknowledgements}

We thank M.~M. Rubio-D\'iez, C. Engelbrecht and M.~M. Kotze for taking additional photometry not included here. We thank S. Mohamed for discussions. PAW, BW and MM acknowledge financial support from the South African NRF and the University of Cape Town. We thank the anonymous referee for a helpful and constructive report. Based on observations made with the SAAO 1.9-m telescope, the Southern African Large Telescope (SALT) under programmes 2012-1-RSA-009 and 2014-2-SCI-060, the VLT at the Paranal Observatory under programme 088.D-0750(A) and the SPM 2.1-m telescope. This work would not have been possible if it were not for the initial discovery of the binary central star of Te~11 enabled by observations made with the Mercator Telescope, operated on the island of La Palma by the Flemish Community, at the Spanish Observatorio del Roque de los Muchachos of the Instituto de Astrof\'isica de Canarias. \textsc{synphot} is a product of the Space Telescope Science Institute, which is operated by the Association of Universities for Research in Astronomy (AURA) for NASA, and \textsc{iraf} is distributed by the National Optical Astronomy Observatory, which is operated by AURA under a cooperative agreement with the National Science Foundation.

\end{document}